\documentclass[graybox]{svmult}

\usepackage{mathptmx}       
\usepackage{helvet}         
\usepackage{courier}        
\usepackage{type1cm}        
\usepackage{makeidx}         
\usepackage{graphicx}        
\usepackage{multicol}        
\usepackage[bottom]{footmisc}

\makeindex             

\begin{document}

\title*{The Planetary Nebulae Populations in the Local Group}
\author{Magda Arnaboldi}
\institute{Magda Arnaboldi \at European Southern Observatory,
  \email{marnabol@eso.org}}
%
%
\maketitle

\abstract*{Planetary nebulae have been used as tracers of light and
  kinematics for the stellar populations in early-type galaxies since
  more than twenty years. Several empirical properties have surfaced:
  for example the invariant bright cut-off of the planetary nebulae
  luminosity function and correlations of the luminosity specific PN
  number with the integrated properties of the parent stellar
  populations. These observed properties are poorly understood in
  terms of a simple model of a ionized nebula expanding around a
  non-evolving central star. In order to make further steps, we need
  to study self-contained systems at know distances whose PN
  populations are sufficiently nearby to permit investigation into
  their physical properties. The galaxies in the Local Group represent
  a valid proxies to study these late phases of evolved stellar
  populations with a spread of metallicities, $\alpha$-element
  enhancements, and star forming histories.}

\abstract{Planetary nebulae have been used as tracers of light and
  kinematics for the stellar populations in early-type galaxies since
  more than twenty years. Several empirical properties have surfaced:
  for example the invariant bright cut-off of the planetary nebulae
  luminosity function and correlations of the luminosity specific PN
  number with the integrated properties of the parent stellar
  populations. These observed properties are poorly understood in
  terms of a simple model of a ionized nebula expanding around an
  non-evolving central star. In order to make further steps, we need
  to study self-contained systems at know distances whose PN
  populations are sufficiently nearby to permit investigation into
  their physical properties. The galaxies in the Local Group represent
  a valid proxies to study these late phases of evolved stellar
  populations with a spread of metallicities, $\alpha$-element
  enhancements, and star forming histories.}

\section{Introduction}
\label{sec:1}
Galactic planetary nebulae (PNs) are roughly 0.3 pc in diameter. Hence
when they are observed in external galaxies at distances larger than
50 Kpc they are identifiable as unresolved emissions of monochromatic
green light at 5007 \AA, that is the strongest optical emission line
of the [OIII] doublet. About 2000 PNs are known in the Milky Way (MW)
out of 200 billion stars and are found mostly in the plane of the
MW. In our own galaxy, 95\% of the stars end their lives as PNs, and
the remaining 5\% as SNs. Single stars can be observed in their PNs
phase because the ionized envelope of a PN is able to re-emit up to 15\% of
the core star UV radiation in one single optical line, the [OIII]
5007\AA\ line, which can be detected in narrow band imaging surveys
against the continuum light emitted by the rest of the stars.  The
total integrated flux in the [OIII] 5007 line emitted by a nebula can
be expressed as
\begin{equation}\label{eq:m}
m_{5007} = -2.5 \log (F_{5007}) - 13.74
\end{equation} 
When these magnitudes are measured for a PN population, they provide
the planetary nebulae luminosity functions (PNLF). In external
galaxies, the PNLFs show several empirical properties, for example the
bright cut-off is observed to be empirically invariant and the overall
shape seems to be reproduced by the Ciardullo's 1989 analytical
formula
\begin{equation} \label{eq:n}
N(M)\propto exp(0.307M) \times (1-exp(3(M*-M)));\,\, M*=-4.51 
\end{equation}

\subsection{The PN.S Key Project on Early-Type Galaxies}
\label{subsec:1.1}
PNs have been used as tracers of light and motions in the outer
regions of early-type galaxies since more than 20 years, starting from
the early models of the mass distribution in M32 by Nolthenius \& Ford
(1986) to the current Planetary Nebulae Spectrograph (PN.S) $-$ Key
Project $-$ for the survey of PNs in early-type galaxies (ETGs). Such
survey includes all ETGs within 20 Mpc, at airmass less than 1.5 which
are observable from the William Herschel telescope on La Palma,
Spain. This survey includes 20 ETGs, from E0 to E4, 11 S0s and it is now
expanded to include spiral galaxies. Observational data are presented
in Coccato et al. (2009) and Cortesi et al. (2013); data products can
be downloaded at {\it
  http://www.strw.leidenuniv.nl/pns/PNS\_public\_web/PN.S\_data.html}

\begin{figure}[b]
\sidecaption
\includegraphics[scale=.40]{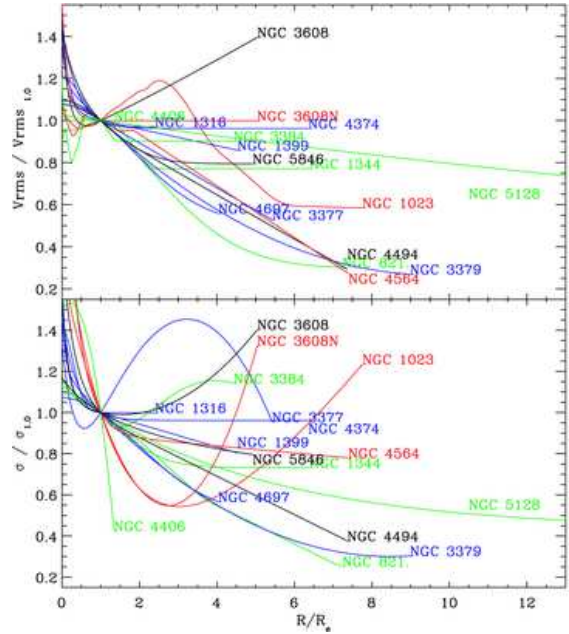}
%
%
\caption{Comparison between radial profiles of $V_{rms}$ (top panel)
  and velocity dispersion (bottom panel) of the sample galaxies,
  obtained by combining the stellar and PNs kinematics along the major
  axis. Profiles have been scaled to the effective radius and
  normalized to their value at $1.0 R_e$. Colors are chosen in order
  to highlight the contrast between lines and thus better distinguish
  different profiles.  The velocity {\rm r.m.s.} and $\sigma/\sigma_0$
  profiles in ETGs as traced by PNs. From the PN.S Key Project,
  Coccato et al. (2009). }
\label{fig:1}       
\end{figure}

The current results of these studies illustrate that the PN number
density follows the surface brightness of the stars that dominate the
light in these galaxies, and whenever PNs and absorption line
kinematics overlap, they agree in all sample galaxies surveyed thus
far. The results from the kinematical studies show that the specific
angular momentum in the outer halo, out to 10 $R_e$, where $R_e$ is
the effective radius of the surface brightness distribution,
correlates with the properties of the high surface brightness
regions, but there are significant deviations (Coccato et al. 2009,
Cortesi et al. 2013). Furthermore, the velocity {\it r.m.s.}
dispersion profiles for the galaxies in this sample show a dichotomy
between flat and quasi Keplerian dispersion profiles, see
Figure~\ref{fig:1}. The question whether there is a continuum or a
separation between the two families is still open, and is going to be
addressed by the new data release from the PN.S key project. Contrary
to the initial claims of quasi Keplerian dispersion profile signaling
lower dark matter content in these systems ( e.g. Romanowsky et
al. 2003), these galaxies can arbor a dark matter halo consistent
with $\Lambda CDM$ and their steep $\sigma$ profiles at large radii
are the results of the R$^{1/4}$ surface brightness profile and strong
radial anisotropy in the outer parts (de Lorenzi et al. 2009, Morganti
et al. 2013).

%

\subsection{PNs in ETGs}
\label{subsec:1.2}

\begin{figure}[b]
\sidecaption
\includegraphics[scale=.40]{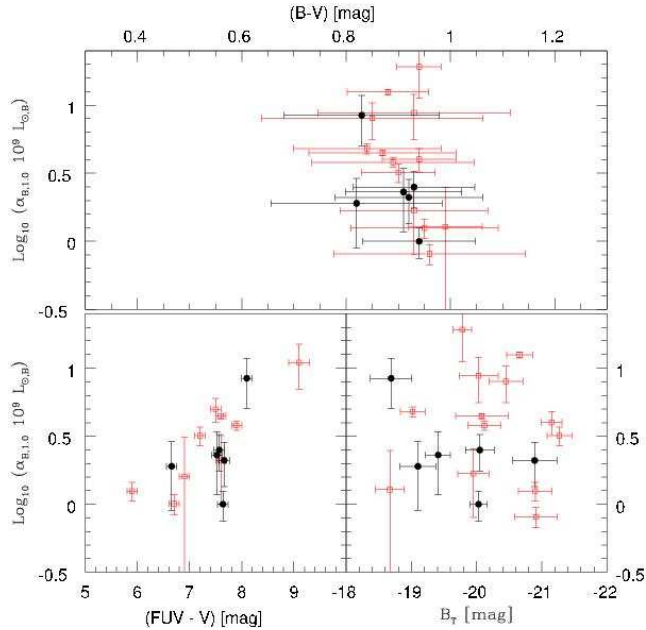}
%
%
\caption{The $\alpha$ parameter values measured in ETGs and
  correlations with $B-V$ color, total B magnitude and $FUV-V$
  color. From Cortesi et al. (2013).}
\label{fig:2}       
\end{figure}

The luminosity specific PN number, or $\alpha$ parameter for short is
given by
\begin{equation}
\alpha = N_{PN} / L_{\odot, Bol} 
\end{equation}
and it is a characteristic number that associates the PN population
with its parent stellar population emitting the bulk of the light
in an extragalactic system. In the simple stellar population theory,
such number is also given by
\begin{equation}
\alpha = N_{PN} / L_{\odot, Bol} = B\tau_{PN}
\end{equation} whether $B$ is the specific evolutionary flux and $\tau_{PN}$ is
the PN visibility lifetime of a PN. In the simple model of a uniformly
expanding nebula around a non-evolving core (Henize \& Westerlund
1963), $\tau_{PN} = D/v_{exp}$ , where $D$ is the diameter of a PN
when its $m_{5007}$ has faded $8$ mags below the bright cut-off, and
$v_{exp}$ is the nebula expansion velocity. For this simple model
$\tau_{PN} = 30000$ yrs. Empirically, the $\alpha$ parameter shows a
constant value in stellar populations with integrated $B-V < 0.8$, and
then the scatter increases for stellar populations with redder colors (Hui
et al. 1993, Ciardullo et al. 2002, Buzzoni et al. 2006, Cortesi et
al. 2013). Specific to ETGs, the $\alpha$ parameter correlates with
$FUV-V$ colors, so that fewer PNs are observed in systems with an
$FUV-V$ color excess, see Figure~\ref{fig:2}.

%
From the relation between $\alpha$ and $\tau_{PN}$, a lower value of
$\alpha$ translates into a shorter visibility lifetime for the PN
population of about $1000 - 5000$ yrs in those systems where the low
values of $\alpha$ are measured.

\section{Surveys of PNs in the Local Group}
\label{sec:2}
In this section, a summary overview of the available surveys for PNs
in the Local Group (LG) is provided, with emphasis on the surveys of the MW
PNs, in the Large Magellanic Cloud (LMC) and the Andromeda galaxy
(M31).

{\it Survey of PNs in the MW} $-$ The PN candidates in the MW are
selected in H$\alpha$ and confirmed spectroscopically. The most recent
survey is the Macquarie/AAO/Strasbourg H$\alpha$ Planetary Nebula
Catalog (MASH I+ MASH II) with $\sim 1200$ PNs detected (Frew et
al. 2013, Parker et al. 2006 (MASH I), and Miszalski et al. 2008 (MASH
II)). These latest results build on and expand the Strasbourg/ESO
catalog of PN ($\sim 2000$ PNs by Acker et al. 1992). The PN
population that is specific to the Bulge includes 802 PNs, with 405
PNs from MASH I + II and 396 PNs from Acker et al. (1992). Independent
surveys of the Bulge PNs (373 PNs) were carried out by Beaulieu et
al. (1999).  Weidmann et al. (2013) also identified PNs in the MW
bulge from VVV survey data (about 381 PNs).

{\it Survey of PNs in the Large Magellanic Cloud} $-$ An extended
survey in the LMC was carried out by Reid \& Parker (2010) in the
central $25$ deg$^2$ area. A total of 740 PNs were detected and
spectroscopically confirmed with the AAT 2dF Spectrograph. The
magnitude range covered by the sample spans 10 mags down the bright
cut-off of the PNLF.

{\it Survey of PNs in M31} $-$ The extensive survey of PNs in M31 was
carried out with the Planetary Nebulae Spectrograph (PN.S) by Merrett
et al. (2006), with the discovery of 2730 PN candidates. For this
sample, the PNLF is complete 4 mag down the brightest cut-off (Merrett
et al. 2006).

{\it Survey of PNs in LG galaxies} $-$ The PN populations in
LG galaxies were surveyed and detected in 16 LG members (Reid
2012). There are no PN detected in LG galaxies with $M_V < -9.8
(\simeq 10^6 L_\odot)$.  Empirically, it is consistent with a maximum
specific PN luminosity number of $\alpha_{max} < 1 PN / 10^7
L_{\odot,Bol}$.

\section{Physical properties of the PN populations - visibility lifetimes and 
planetary nebulae luminosity functions}
\label{sec:3}
{\it MW Bulge} - The distribution of angular diameter in arcsec for
Bulge and Disk PNs is available from the MASH-{\rm I + II} survey,
with 90\% of the whole population within the maximum angular diameter
of 35 arcsec. In the case of the PN sample restricted to the MW Bulge,
the plot $\log (SB)$ vs. $\log(D)$ shows that round PNs are the
brightest and most compact objects, with average diameter of 0.3 pc:
clearly these PNs are those preferentially detected when PN
populations are observed in external galaxies.

There are 560 PNs in the Galactic Bulge with average diameter about
0.3 pc. The PN visibility lifetime $\tau_{PN}$ for this volume limited
sample can be estimated using $v_{exp}$ and $D_{PN}$, as $D_{PN} /
v_{exp}$. Given that the average expansion velocity of Bulge PNs is
$30$kms$^{-1}$, thus the observable lifetime of a Galactic Bulge PN
is then only a few $10^3$ years, which is consistent with visibility
lifetime of PN samples in ETGs.

Kovacevic et al. (2011) carried out the [OIII] narrow band imaging of
the PN population in the central $10^\circ \times 10^\circ$ region
towards the MW Bulge. This sample accounts for 80\% of known PNs in
this region. Fluxes and diameters are measured from narrow band imaging
with the MOSAIC-II camera on the 4-m Blanco Telescope at the
CTIO. They surveyed $~60 deg^2$ uniformly with a narrow band filter
centered on the [O III] $\lambda$ 5007 line; 104 objects were used for
calibration of the $m_{5007}$ magnitudes of 228 PNs. At 8 kpc, $m*$ is
10.00 ($M*=-4.51$), and one can ask whether the PN population
associated with an old $\alpha-$element enhanced parent stellar
population follows the PNLF analytical formula suggested by Ciardullo
et al. (1989). The empirical PNLF for the MW Bulge is steeper than
what is described by the Ciardullo's 1989 analytical formula, in a way
similar to what is observed in the halo of M87 (Longobardi et
al. 2013). For more detail, we refer to Arnaboldi, Longobardi,
Gerhard in prep.

{\it LMC } - Reid \& Parker (2010) build the PNLF for the extended
sample of PNs in the LMC. In order to use the PNLF for distance
determination, they corrected the [OIII] fluxes from the confirmed
spectra for line-of-sight reddening only. While the bright cut-off is
consistent with M*=-4.51, once corrected for the metallicity
dependence, the Ciardullo's 1989 formula is a poor fit to the
empirical PNLF. The major departure from the analytical formula occurs
1.5 mags from the bright cut-off and lasts for 1.5 mag. The empirical
PNLF is flatter in these magnitude ranges than what predicted by the
double exponential formula (eq. 2). This very sensitive survey
illustrates another new feature of the PNLF: a peak in the PN number
density which is caused by a rapid rise 4 mags below the brightest
PN. This feature is characteristics of models where the [OIII] 5007
\AA\ emission is driven by the rapid evolution of the central
star. The peak and the steep descent represents the
point at which the central star starts its trajectory down towards the
white dwarf cooling track. Such a peak in the PNLF is predicted by the
models of Me\'ndez et al. (1993)

The very deep survey of PNs in the LMC shows that Ciardullo's
analytical formula fails to reproduce the faint end of the PNLF, with
candidates being detected up to 2 mags fainter than the expected limit
of 8 mag down the brightest PNs.

{\it M31} - The PN population in M31 illustrates the strength of PNs as
tracers of light and motions of the stars that dominate the bulk of
the light emitted in this galaxy. As shown by Merrett et al. (2006),
the PN number density profile follows the surface brightness profile
along the major and the minor axis out to 2 and 1 degree,
respectively, from the galaxy center. The PNLF for the entire
population is complete 4 magnitudes down the bright cut-off and it is
well fitted by eq.~\ref{eq:n}. Deviations from the Ciardullo's
analytical formula are measured for the PNLF in the
very central bright region by Pastorello et al. (2013). The kinematics
of the PN sample shows the presence of a substructure (Merrett et
al. 2003), while the bulk of the PN follow the stellar disk, with a
clear signature for the asymmetric drift and a flattening of
$\sigma_v$ at large radii in the disk, which is signaling an increase
in the DM content.

\begin{figure}[b]
\sidecaption
\includegraphics[scale=.50]{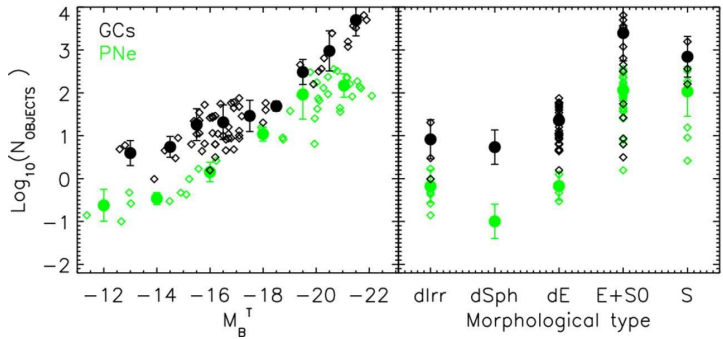}
%
%
\caption{Expected number of sources (GCs or PNs) as a function of
  galaxy absolute magnitude (left) and morphology (right). The
  expected number of GCs and PNs is computed from: (i) the specific
  frequencies of GCs and PNs and (ii) the total galaxies’ luminosities
  in the V and B bands, for GCs and PNs, respectively. The observed
  number of PNs approaches the theoretical predictions from
  single-stellar-population models (Buzzoni et al. 2006) for LMC-like
  and L* systems. From Coccato et al. (2013).}
\label{fig:3}       
\end{figure}

{\it LG galaxies} - The surveys of PN population in the LG
galaxies can provide useful information on the frequency of PNs in
different systems, and whether there are systems with higher specific
frequencies as function of luminosities and morphological types. These
analysis were carried out by Reid (2012) and Coccato et al. (2013);
they show that galaxy systems with luminosities less than
$10^8$L$_{\odot, Bol}$ have fewer PNs that the number predicted from
the theoretical maximum luminosity specific PN number of $1 PN/1.5
\times 10^6$ L$_\odot$ (Buzzoni et al. 2006). The observed number of
PNs approaches the theoretical predictions for LMC-like and L* systems,
see Figure~\ref{fig:3}.

%

\section{Physical properties of the PN populations in old-metal rich vs. poor-star forming stellar populations.}
\label{sec:4}
The oldest and most metal rich stellar populations are found in the
bright, red, and passively evolving galaxies at the centers of massive
clusters. The nearest such object is the galaxy M87, in the core of
the Virgo cluster. Narrow band imaging surveys by Jacoby et al. (1990)
provided the first PN candidates in M87, and the first PNs associated
with the intracluster light (ICL) were discovered by Arnaboldi et
al. (1996) and Feldmeier et al. (1998). The spectroscopic confirmation
based on the detection of the [OIII] 4959/5007 \AA\ doublet in single
objects was obtained by Arnaboldi et al. (2004) and Doherty et
al. (2009). These 52 PNs turned out to be associated to the halo of
M87 and ICL, as indicated by their line-of-sight velocities.

An important step forward in the study of PNs in M87 and the Virgo
cluster comes with the availability of SuprimeCAM on the Subaru
telescope, that allows the complete coverage of 0.25 deg$^2$ FoV, with
the collecting power of an 8 meter telescope. The results for the PN
populations in M87 and ICL are reported in Longobardi et al. (2013)
and Arnaboldi et al. (2003).

The deep narrow band imaging survey carried out by Longobardi et
al. (2013) led to the identification of 688 PN candidates in the outer
regions of M87, covering a radial range from 10 to 150 kpc. This
sample is corrected for the contamination by Ly$\alpha$ emitters at
redshift 3.14, [OII] emitters at redshift 0.34, faint continuum stars,
and for color and spatial incompleteness. Longobardi et al. (2013)
built the PN number density profile and compared it with the surface
brightness profile of M87 in the V band. Deviations were clearly
detected with the number density profile being flatter than the
surface brightness profile at large radii. Such behavior was
reproduced by a two component model where the PNs in the survey field
come from the PN population associated with the M87 halo and the
ICL. A two component model reproduces the observed profile when
$\alpha_{2.5, ICL} = 3 \times \alpha_{2.5, M87}$. The values of
$\alpha$ translate into PN visibility lifetimes that are $\tau_{PN}=
1.4 \times 10^4$ for PNs in the ICL and $4.5 \times 10^3$ for PNs in
the M87 halo.

\begin{figure}[b]
\sidecaption
\includegraphics[scale=.80]{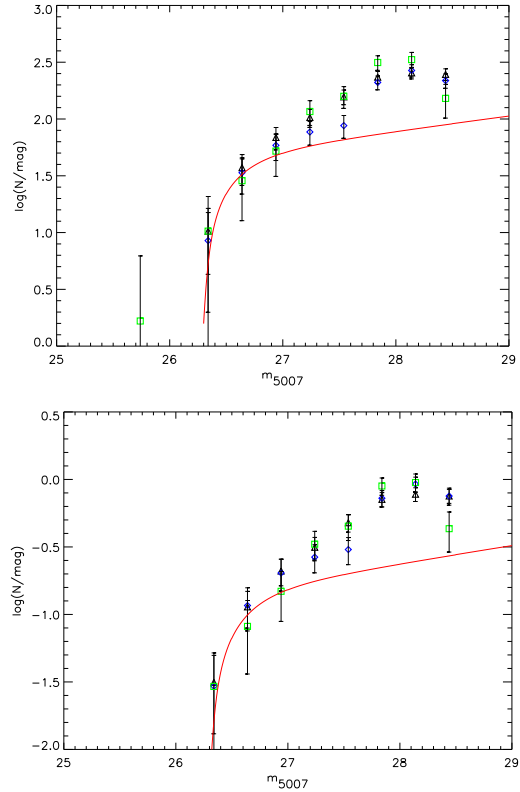}
%
%
\caption{ Top panel: empirical PNLFs in three radial ranges corrected
  for color and detection incompleteness: PN candidates within $6.5'$
  from M 87 center – triangles, PN candidates between $6.5'$ and $13.5'$
  from M 87 center – diamonds, PN candidates in the outermost region
  (distances greater than $13.5'$) – squares. The respective Ly$\alpha$
  contribution expected in each radial bin was subtracted. Magnitudes
  are binned in 0.3 mag bins and the error bars represent the
  1$\sigma$ uncertainty from counting statistics combined with the
  uncertainty from cosmic variance in the Lyα density. The red solid
  line is the convolved analytical formula of Ciardullo et al. (1989)
  for distance modulus 30.8. Lower panel: same as for the upper plot,
  but now the three PNLFs are normalized at the total number of
  objects in each radial bin. The three data sets are consistent with
  being drawn from the same underlying distribution. From Longobardi
  et al. (2013).}
\label{fig:4}       
\end{figure}

Longobardi et al. (2013) used their large PN sample in the outer
regions of M87 to investigate the properties of the PNLF. They
compared it with the PNLF predicted from Ciardullo's 1989 analytical
formula and a distance modulus of 30.8. They reported strong
deviations at 1.5 mag below the brightest cut-off with the observed
gradient being steeper than what is predicted by the analytical formula
in this magnitude range. Longobardi et al. (2013) compared the PNLFs
extracted in three radial bins, and showed that the three PNLFs
have high Kolmogorov-Smirnov (KS) probabilities of being drawn from
the same distribution, while the KS test rejects the Ciardullo's
formula, see Figure~\ref{fig:4}.

%

\begin{figure}[b]
\sidecaption
\includegraphics[scale=.80]{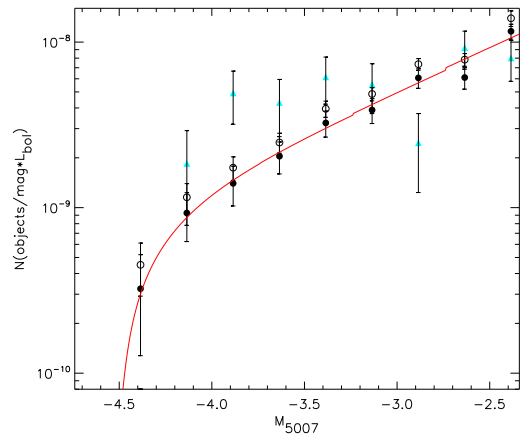}
%
%
\caption{Luminosity-normalized PNLFs for the bulge of M31 (blue triangles)
  and the halo of M87 (black circles). Data are binned into 0.25 mag
  intervals. The red line shows the improved fit with the second
  parameter $c_2$, in addition to the distance modulus and the
  normalization factor $c_1$. The M87 PNLF has a steeper slope towards
  faint magnitudes than M31. From Longobardi et al. (2013).}
\label{fig:5}       
\end{figure}

The results are significant for old stellar populations because the
steepening of the PNLF is thus shown to be present in all three radial
bins, while it is known that the ICL contributes mostly at the largest
radius. We can then compare the observed properties of the PNLF in M87
with the PNLF in the M31 bulge, which was used to calibrate the
analytical formula from Ciardullo et al. (1989). The PN sample for M31
and the halo of M87 are normalized by the sampled luminosity and then
corrected for the distance modulus. In Figure~\ref{fig:5}, the results
show that within one mag of the bright cutoff, the M87 population has
fewer PNs than the PNLF of M31, i.e. the slope of the PNLF for the
halo of M87 is steeper towards fainter magnitudes.

In old stellar populations, one expects the PN central stars to be
mostly low mass cores, with $M_{core} \leq 0.55$M$_\odot$. Thus, the
slope of the PNLF may turned out steeper than the slope predicted by
the fading of a uniformly expanding sphere ionized by a non-evolving
star (Henize \& Westerlund 1963). The comparison between the PNLFs of
M87 and M31 may indicate that the M87 halo hosts a stellar population
with a larger fraction of low mass cores, with respect to the M31
bulge. Comparison of the properties of PNLFs and their gradients 1.5
mag below the brightest PNs is further explored in Arnaboldi,
Longobardi, Gerhard in prep.

\section{Conclusions}
\label{sec:5}

The recent work on the PN populations in the halo of M87 and ICL shows
that a second parameter is needed to generalize the Ciardullo's 1989
formula and account for population effects. In addition to the distance
modulus and the normalization coefficient $c_1$, we need to vary the
slope $c_2$, which is related to the gradient of the PNLF at
magnitudes fainter then the PNLF cut-off.

Furthermore, one concludes 1) that PN populations are ubiquitous - deep
photometry and kinematics from PNs show that they trace the bulk of
star light in LG galaxies, early-type galaxies and in
extended luminous halos, 2) that PN populations and their PNLFs show
dependencies on stellar populations which are quantified by the
gradient in the PNLF at magnitudes below the bright cut-off, and 3)
that M* is empirically observed to be invariant in the systems studied
thus far.

%

\begin{acknowledgement}
MAR would like to thank her collaborators for their enthusiasm and
support, in particular Alessia Longobardi, Ortwin Gerhard and Lodo
Coccato. MAR would like to thank the Scientific Organizing Committee
for the invitation to give this review and the opportunity to take
part in this very special and vibrant conference. MAR acknowledges ESO
for financial support. All the best to Bruce and David!
\end{acknowledgement}

\section{References}
Acker, A., Marcout, J., Ochenbein, F. et al. 1992, Strabourg-ESO Catalogue of 
Galactic Planetary Nebulae (Garching:ESO)\\
Arnaboldi, M., Freeman, K., Mendez, R., et al. 1996, ApJ, 472, 175\\
Arnaboldi, M., Freeman, K., Okamura, S., et al. 2003, AJ, 125, 514\\
Arnaboldi, M., Gerhard, O., Aguerri, J. , et al. 2004, ApJL, 614, 33\\ 
Beaulieu, S., Dopita, M., Freeman, K. 1999, ApJ, 515, 610\\
Buzzoni, A., Arnaboldi, M., Corradi, R. 2006, MNRAS, 368, 887\\
Ciardullo, R., Jacoby, G., Ford, H. et al. 1989, ApJ, 339, 53\\  
Ciardullo, R., Feldmeier, J., Jacoby, G. 2002, ApJ, 577, 31\\
Coccato, L., Arnaboldi, M., Gerhard, O. 2013, MNRAS, 436, 1322\\
Coccato, L., Gerhard, O., Arnaboldi, M. et al. 2009, MNRAS, 394, 1249\\
Cortesi, A., Arnaboldi, M., Coccato, L. et al. 2013, A\&A, 549, 115\\
Doherty, M., Arnaboldi, M., Das, P., et al. 2009, A\&A, 502, 771\\ 
de Lorenzi, F. Gerhard, O., Coccato, L., et al. 2009, MNRAS, 395, 76\\
Feldmeier, J., Ciardullo, R., Jacoby, G. 1998, ApJ, 503, 109\\ 
Frew, D., Boji\^ci\'c, I., Parker, Q. 2013, MNRAS, 431, 2\\
Henize, K. \& Westerlund, B. 1963, ApJ, 1963, 137, 747\\
Hui, X. Ford, H., Ciardullo, R., et al. 1993, ApJ, 414, 463\\ 
Longobardi, A., Arnaboldi, M., Gerhard, O., et al. 2013, A\&A, 558, 42\\
Jacoby, G., Ciardullo, R., Ford, H. 1990, ApJ, 356, 332\\
Kovacevic, A., Parker, Q., Jacoby, G., et al. 2011, MNRAS, 414, 860\\
Mendez, R.H., Kudritzki, R., Ciardullo, R., et al. 1993, A\&A, 275, 534\\
Merrett, H., Kuijken, K., Merrifield, M., et al. 2003, MNRAS, 346, 62\\ 
Merrett, H., Merrifield, M., Douglas, N., et al. 2006, MNRAS, 369, 120\\ 
Miszalski, B., Parker, Q., Acker, A.,  et al. 2008, MNRAS, 384, 525 (MASH II)\\
Morganti, L., Gerhard, O., Coccato, L., et al. 2013, MNRAS, 431, 357\\
Nolthenius, R. \& Ford, H. 1986, ApJ, 305, 600\\
Parker, Q., Acker, A., Frew, D., et al. 2006, MNRAS, 373, 79 (MASH I)\\ 
Pastorello, N., Sarzi, M., Cappellari, M., et al. 2013, MNRAS, 430, 121\\
Reid, W. 2012, IAU Symp., 283, 227\\
Reid, W \& Parker, Q. 2010, MNRAS, 405, 1349\\
Romanowsky, A., Douglas, N. Arnaboldi, M., et al. 2003, Science, 301, 1696\\
Weidmann, W.,  Gamen, R., van Hoof, P., et al. 2013, A\&A, 552, 74\\

\end{document}